\documentclass[11pt]{article}
\usepackage{graphicx}
\begin{document}

\begin{center}

\vbox{\vspace{20mm}}

{\LARGE \bf Poincar\'e Sphere and a Unified Picture of \\Wigner's Little Groups}\\
 \vspace{6mm}
Y. S. Kim \\
Department of Physics, University of Maryland, \\
College Park, Maryland 20742, U.S.A.\\
\vspace{10mm}
\end{center}

\abstract{It is noted that the Poincar\'e sphere for polarization optics
contains the symmetries of the Lorentz group.  The sphere is thus capable
of describing the internal space-time symmetries dictated by Wigner's
little groups.  For massive particles, the little group is like the
three-dimensional rotation group, while it is like the two-dimensional
Euclidean group for massless particles.  It is shown that the Poincar\'e
sphere, in addition, has a symmetry parameter corresponding to reducing
the particle mass from a positive value to zero.  The Poincar\'e sphere
thus the gives one unified picture of Wigner's little groups for massive
and massless particles.}

\vspace{50mm}
\noindent This paper is based on a plenary talk delivered at the
Wigner Research Symposium entitled ``Wigner 111 - Colourful and Deep''
(Budapest, Hungary, November 2013), to be published in the Proceedings.

\newpage

\section{Introduction}\label{intro}
In 1939~\cite{wig39}, Wigner considered the subgroups of the Lorentz group
whose transformations leave the four-momentum of a given particle invariant.
These subgroups therefore dictate the internal space-time symmetries of
the particles.  They are called Wigner's little groups. The little groups
for massive and massless particles are like $O(3)$ and $E(2)$
respectively.
\par
The physics of $O(3)$-like little group is transparent.  For a massive
particle at rest, it can have the spin whose direction can be rotated
according to the three-dimensional rotation group.  On the other hand,
the $E(2)$-like group for massless particles has a stormy history, and
its physics had not been settled until 1990~\cite{kiwi90jmp}. It is not
difficult to see its rotational degree of freedom corresponds to the
helicity.  The two translation-like variable collapse into that of one
gauge-transformation variable.
\par
The question is whether it is possible to derive this $E(2)$-like little
group as a limiting case of the $O(3)$-like little group massive particles.
It is not possible to reduce the value of mass continuously to zero within
the framework of the Lorentz group because the mass is an invariant quantity.
However, it is possible to make the momentum infinite where the
mass-hyperbola becomes tangent to the light cone, and then reduce the
energy to a finite value along this light cone~\cite{kiwi90jmp}.
\par
It was noted that the Poincar\'e sphere is based on the two-by-two
coherency matrix consisting of four stokes parameters constructed from the
two-component Jones vector~\cite{born80,bros98}.  The Jones vector consists
of the two transverse components of a light beam.  The transformation
applicable to this vector is that of $SL(2,c)$ isomorphic to the Lorentz
group.   Thus, the Poincar\'e sphere contains the symmetries of the Lorentz
group~\cite{kn13symm}.
\par
In addition, the coherency matrix measures the coherence or decoherence
between the two transverse electric fields~\cite{hkn97,saleh07}.  The
determinant of this two-by-two matrix gives the degree of coherence.   It
is shown in this note that the variation of this degree corresponds to
variation of the mass variable not allowed in the Lorentz group.  Thus,
the Poincar\'e sphere allows to reduce the mass continuously to zero from
a positive value.

\par
In Sec.~\ref{lorentz}, we present the two-by-two representation of the
Lorentz group where both the space-time and momentum-energy four-vectors
are written in the form  of the two-by-two matrix.   The determinant of
the momentum matrix is $(mass)^2$.  In Sec.~\ref{wigner}, Wigner's little
group is formulated in the language of two-by-two matrices as in the case
of the coherency matrix.
In Sec.~\ref{poincs}, it is noted that the coherency matrix for
the Poincar\'e sphere contains the symmetries of the Lorentz group.  In
addition, it is shown the degree of coherence corresponds to the particle
mass, which can be changed continuously from a positive value to zero.

\section{Two-by-two representation of the Lorentz Group}\label{lorentz}
Let us start with the two-by-two matrix
\begin{equation}\label{2b2}
X = \pmatrix{ t + z  &  x - iy \cr x + iy & t - z} ,
\end{equation}
Then its determinant is $\left(t^2 - z^2 - x^2 -y^2\right)$.  Thus,
the Lorentz transformation is a determinant-preserving transformation.
We can the consider a unimodular matrix of the form~\cite{naimark54}
\begin{equation}\label{alphabeta}
 G = \pmatrix{\alpha & \beta \cr \gamma & \delta},
  \quad\mbox{and}\quad
  G^{\dagger} =
  \pmatrix{\alpha^* & \gamma^* \cr \beta^* & \delta^*} ,
\end{equation}
with
\begin{equation}\label{detone}
    \det{(G)} = 1 ,
\end{equation}
and the transformation
\begin{equation}\label{naim}
X' = G X G^{\dagger} .
\end{equation}
This can be explicitly written as
\begin{equation}\label{lt01}
\pmatrix{t' + z' & x' - iy' \cr x + iy & t' - z'}
 = \pmatrix{\alpha & \beta \cr \gamma & \delta}
  \pmatrix{t + z & x - iy \cr x + iy & t - z}
  \pmatrix{\alpha^* & \gamma^* \cr \beta^* & \delta^*} .
\end{equation}
Since $G$ is not a unitary matrix, Eq.(\ref{naim}) not a unitary
transformation.  This two-by-two transformation can be translated into
the language of the conventional four-by-four representation of the
Lorentz group~\cite{hkn97,knp86}.
\par

The energy-momentum four-vector can also be written as a two-by-two matrix.
It can be written as
\begin{equation}\label{mom00}
P = \pmatrix{p_0 + p_z & p_x - ip_y \cr
p_x + ip_y & p_0 - p_z} ,
\end{equation}
with
\begin{equation}
\det{(P)} = p_0^2 - p_x^2 - p_y^2 - p_z^2,
\end{equation}
which means
\begin{equation}\label{mass}
\det{(P)} = m^2,
\end{equation}
where $m$ is the particle mass.
\par
The Lorentz transformation can be written explicitly as
 \begin{equation}
 P' = G P G^{\dagger} .
 \end{equation}
This is an unimodular transformation, and the mass is a Lorentz-invariant
variable.
\par
If the elements of the $G$ matrix are complex, but it has six independent
parameters due to the condition of Eq.(\ref{detone}), as in the case of
the Lorentz group.

Furthermore, the elements of the $X$ and $P$ matrices have to be real.
This means that the $y$ components have to vanish.  The transformation
is applicable only to the $t, z,$ and $x$ coordinates.

In mathematics, the group represented by the two-by-two matrix of
Eq.(\ref{alphabeta}) is called $SL(2,c)$ locally isomorphic to the group
of  Lorentz transformation matrices applicable to the four-dimensional
Minkowskian space.  The three-parameter subgroup consisting only of real
matrices is called $Sp(2)$ locally isomorphic to the group of
three-by-three matrices performing Lorentz transformations on the
space of $t, z$, and $x$.

\par
For a physical system invariant under rotations around the $z$ axis,
it is enough to study the $Sp(2)$ subgroup.  Here, for all practical
purposes, it is sufficient to work with the following three matrices.
\begin{eqnarray}\label{mat12}
&{}& R(\theta)= \pmatrix{ \cos(\theta/2) &  -\sin(\theta/2) \cr
                  \sin(\theta/2) & \cos(\theta/2)},  \nonumber\\[1.5ex]
&{}& B(\eta)= \pmatrix{ e^{\eta/2} & 0 \cr 0 & e^{-\eta/2}},
\qquad
S(\lambda) = \pmatrix{ \cosh(\lambda/2) &  \sinh(\lambda/2) \cr
                  \sinh(\lambda/2) & \cosh(\lambda/2)} ,
\end{eqnarray}
corresponding to the rotation around the $y$ axis, and the boost along
the $z$ and $x$ directions, respectively. The rotation around the $z$ axis
corresponds to
\begin{equation}\label{rotz}
 Z(\phi) = \pmatrix{ e^{i\phi/2} & 0 \cr 0 & e^{-i\phi/2}
 } .
\end{equation}

\section{Internal Space-time Symmetries}\label{wigner}
In 1939~\cite{wig39}, Wigner considered the subgroups of the Lorentz group
whose transformations leave the four-momentum of a given particle invariant.
We shall use the word "Wigner matrix" which leaves the four-momentum
invariant.  Then the Wigner matrix $W$ is defined as
\begin{equation}
P = W P W^{\dagger} .
\end{equation}
The four-momentum can be brought to the form proportional to
\begin{equation}
\pmatrix{ 1 & 0 \cr 0 & 1 } , \quad\mbox{or}\quad
\pmatrix{ 1 & 0 \cr 0 & -1 } ,
\end{equation}
if the mass of the particle is positive or imaginary respectively.  If
the mass is positive, it can be brought to its rest frame with zero momentum.
If the mass is imaginary, it can be bought to the frame with zero energy
and momentum along the $z$ direction.  The Wigner matrices which leave
the these four-momenta are $R(\theta)$ and $S(\lambda)$ of Eq.(\ref{mat12})
respectively.

\par
If the particle is massless, the four-momentum
becomes proportional to
\begin{equation}
 \pmatrix{1 & 0 \cr 0 & 0 },
\end{equation}
if the momentum is along the $z$ direction.  Its Wigner matrix takes the
form
\begin{equation}\label{trian}
\pmatrix{1 & -\gamma \cr 0 & 1}.
\end{equation}
While the physics of the Wigner matrix is transparent for massive
and imaginary-mass particles, this triangular matrix is strange.
If it is translated into the four-by-four matrix, the transformation
matrix becomes
\begin{equation}\label{e44}
\Gamma(\gamma, \phi) = \pmatrix{1 + \gamma^2/2  & - \gamma^2/2
  & \gamma  & 0 \cr
 \gamma^2/2  &  1 - \gamma^2/2  & \gamma & 0 \cr
 -\gamma & \gamma  & 1 & 0 \cr
  0 &  0  & 0 & 1}
\end{equation}
applicable to the $ (t, z, x, y)$ coordinates.  This matrix is
in Wigner's original paper but has a stormy history.  It was later
shown that this matrix performs a gauge transformation when applied
to the photon four-potential~\cite{kiwi90jmp}.

Our next question is whether it is possible to obtain the Wigner matrix
of Eq.(\ref{trian}) starting from $R(\theta)$ of Eq.(\ref{mat12}) for
the massive particle.  It is not possible to change the mass within the
frame of the Lorentz group.  However, let us look at the matrix
$B(\eta)$ of Eq.(\ref{mat12}).  This matrix corresponds to a Lorentz
boost which will transform the momentum of the massive particle from
zero to $p$ along the $z$ direction, with the parameter $\eta$ defined
as
\begin{equation}
e^{\eta} = \left[\frac{p_0 + p}{p_0 - p}\right]^{1/2} ,
\end{equation}
which becomes $2p/m,$ for large values of the momentum. This limit can
also be obtained for small values of $m$, even though the variation of
this parameter is not allowed in within the framework of the Lorentz group.

\par
\begin{figure}
\centerline{\includegraphics[scale=0.7]{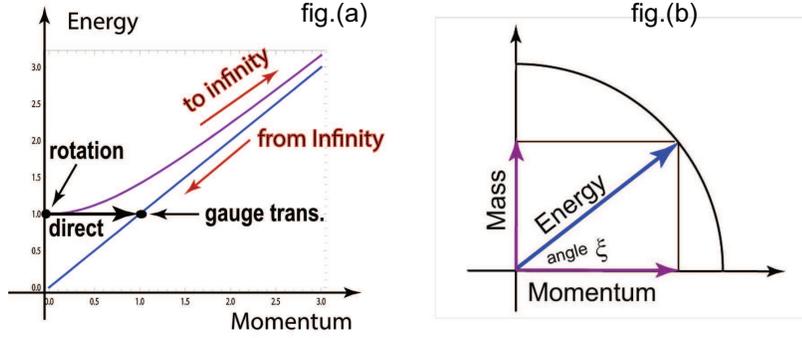}}
\caption{Transition from the $O(3)$-like little group for a massive
particle to the $E(2)$-like little group.  Within the framework of the
Lorentz group, the particle mass is a Lorentz invariant quantity.  However
all mass hyperbolas collapse into the light cone when the particle momentum
becomes infinity.  Thus, it is possible to go to infinity first and come
back after jumping to the light cone, as described in fig.(a).  This figure
also tells how to go to the massless case directly within the symmetry
available from the Poincar\'e sphere.  As is illustrated in fig.(b),
it is possible have a symmetry where the mass and momentum obey the
triangular rule for a fixed energy in the symmetry offered by the
Poincar\'e sphere.}\label{mzero}
\end{figure}

Let us apply this Lorentz boost to the Wigner matrix $B(\theta)$ for the
massive particle.  Then
\begin{equation}
B(\eta) R(\theta) B^{-1}(\eta) =
\pmatrix{\cos(\theta/2) & - e^{\eta}\sin(\theta/2) \cr
        e^{-\eta} \sin(\theta/2) & \cos(\theta/2) } .
\end{equation}
We can make $\eta$ very large and make $\theta$ very small, so that
$e^{\eta}\sin(\theta/2)$ remains as a finite number $\gamma$.  Then
the above expression becomes the Wigner matrix of Eq.(\ref{trian}).
This process is illustrated in Fig.~\ref{mzero}(b).

\section{Symmetries derivable from the Poincar\'e Sphere}\label{poincs}
The Poincar\'s sphere was originally developed as an instrument for
studying polarization of light waves~\cite{born80, bros98}.  If the
light wave propagates along the $z$ direction, its electric field is
perpendicular to the momentum.  It can have its $x$ and $y$ component.
Let us start with the Jones vector defined as
\begin{equation}\label{jvec01}
\pmatrix{\psi_1(z,t) \cr \psi_2(z,t)} =
\pmatrix{ \exp{[i(kz - \omega t)]}  \cr \exp{[i(kz - \omega t)] }
} ,
\end{equation}
where the upper and lower component of this column vector and
$x$ and $y$ components of the electric field.  We start here with
the case with two identical components.
\par
To this vector, we can apply the phase-shift matrix of the form
\begin{equation}\label{rotz22}
\pmatrix{ e^{i\phi/2} & 0 \cr 0 & e^{-i\phi/2} }.
\end{equation}
This matrix has the same mathematical expression as the rotation matrix
given in Eq.(\ref{rotz}), but performs a different physical transformation
in polarization optics.

\par
We can apply the attenuation matrix of the form of $B(\eta)$ given in
Eq.(\ref{mat12}), and the matrix of the form of $R(\theta)$ in order to
rotate the polarization coordinate around the $z$ direction. Then, repeated
applications of these four matrices will lead to the most general form of
$SL(2,c)$ matrix given in Eq.(\ref{alphabeta}), and the application of this
matrix will lead to the most general form of the Jones vector.
These two-by-two matrices perform two different transformations in physics,
as shown in Table~\ref{tab33}.

\begin{table}
\caption{Polarization optics and special relativity sharing the same set
of matrices.  Each matrix has its well-defined role in optics and
relativity.  The determinant of the Stokes or the four-momentum matrix
remains invariant under $SL(2,c)$ transformations.  This determinant is
the $(mass)^2$ in particle physics, while it corresponds to the decoherence
parameter in optics.  In particle physics, the determinant cannot be
changed, while it is a variable that can be changed.}\label{tab33}
\vspace{2mm}
\begin{center}
\begin{tabular}{llcll}
\hline
\hline \\[0.5ex]
Polarization Optics &\hspace{10mm}& Transformation Matrix &\hspace{10mm} &
 Particle Symmetry \\[1.0ex]
\hline \\
Phase shift by $\phi$  &{}&
$\pmatrix{e^{\phi/2} & 0\cr 0 & e^{-i\phi/2}}$
&{}&  Rotation around $z$.
\\[4ex]
Rotation around $z$  &{}&
$\pmatrix{\cos(\theta/2) & -\sin(\theta/2)\cr
    \sin(\theta/2) & \cos(\theta/2)}$
&{}&  Rotation around  $y$.
\\[4ex]
Squeeze along $x$ and $y$  &{}&
$\pmatrix{e^{\eta/2} & 0\cr 0 & e^{-\eta/2}}$
&{}&  Boost along $z$.
\\[4ex]
Squeeze along $45^o$  &{}&
$\pmatrix{\cosh(\lambda/2) & \sinh(\lambda/2)\cr \sinh(\lambda/2)
                & \cosh(\lambda/2)} $
&{}&   Boost along $x$.
\\[4ex]
 $(\sin\xi)^2$ &{}& Determinant &{}&  (mass)$^2$
\\[4ex]
\hline
\hline\\[-0.8ex]
\end{tabular}
\end{center}
\end{table}

\par

However, the Jones vector  cannot tell us whether the two components
are coherent with each other.  In order to address this important degree of
freedom, we use the coherency matrix defined as~\cite{born80,saleh07}
\begin{equation}\label{cocy11}
C = \pmatrix{S_{11} & S_{12} \cr S_{21} & S_{22}},
\end{equation}
where
\begin{equation}\label{cocy12}
<\psi_{i}^* \psi_{j}> = \frac{1}{T}
\int_{0}^{T}\psi_{i}^* (t + \tau) \psi_{j}(t) dt,
\end{equation}
where $T$ is a sufficiently long time interval.
Then, those four elements become~\cite{hkn97}
\begin{eqnarray} \label{cocy15}
&{}& S_{11} = <\psi_{1}^{*}\psi_{1}> = 1 , \qquad
S_{12} = <\psi_{1}^{*}\psi_{2}> = (\cos\xi)e^{-i\phi} , \nonumber \\[1ex]
&{}& S_{21} = <\psi_{2}^{*}\psi_{1}> = (\cos\xi)e^{+i\phi} ,  \qquad
S_{22} = <\psi_{2}^{*}\psi_{2}>  = 1 .
\end{eqnarray}
The diagonal elements are the absolute values of $\psi_1$ and $\psi_2$
respectively.  The angle $\phi$ could be different from the value of
the phase-shift angle given in Eq.(\ref{rotz22}), but this difference
does not play any role in our reasoning.  The off-diagonal elements could
be smaller than the product of $\psi_1$ and $\psi_2$, if the two
polarizations are not completely
coherent.
\par
The angle $\xi$ specifies the degree of coherency.  If it is zero, the
system is fully coherent, while the system is totally incoherent if $\xi$
is $90^o$.   This can therefore be called the ``decoherence angle.''
While the most general form of the transformation applicable to the
Jones vector is $G$ of Eq.(\ref{alphabeta}), the transformation
applicable to the coherency matrix is
\begin{equation}\label{cocy17}
           C' = G~C~G^{\dagger} .
\end{equation}
The determinant of the coherency matrix is invariant
under this transformation, and it is
\begin{equation}
\det(C) =  (\sin\xi)^2 .
\end{equation}
Thus, angle $\xi$  remains invariant.  In the language
of the Lorentz transformation applicable to the four-vector, the
determinant is equivalent to the $(mass)^2$ and is therefore a
Lorentz-invariant quantity.

\par
The coherency matrix of Eq.(\ref{cocy11}) can be diagonalized to
\begin{equation}\label{cocy66}
 \pmatrix{1 + \cos\xi & 0 \cr 0 & 1 - \cos\xi}
\end{equation}
by a rotation.  Let us then go back to the four-momentum matrix of
Eq.(\ref{mom00}).  If $p_x = p_y = 0$, and $p_z = p_{0} \cos\xi$ ,
we can write this matrix as
\begin{equation}\label{cocy68}
p_0 \pmatrix{1 + \cos\xi & 0 \cr 0 & 1 - \cos\xi} .
\end{equation}
\par
Thus, with this extra variable, it is possible to study the little groups
for variable masses, including the small-mass limit and the zero-mass case.
For a fixed value of $P_0$, the $(mass)^2$ becomes
\begin{equation}\label{cocy70}
  (mass)^2 = \left(p_{0}\sin\xi\right)^2 , \quad\mbox{and}\quad
  (momentum)^2 = \left(p_{0}\cos\xi\right)^2 ,
\end{equation}
resulting in
\begin{equation}\label{cocy75}
(energy)^2 = (mass)^2 + (momentum)^2.
\end{equation}
\par
This transition is illustrated in Fig.~\ref{mzero}.  We are interested in
reaching a point on the light cone from a mass hyperbola while keeping
to the energy fixed~\cite{kn13symm}.   According to this figure,
we do not have to make an excursion to infinite-momentum limit.  If the
energy is fixed during this process, Eq.(\ref{cocy75}) tells the mass
and momentum relation, and Fig.~\ref{mzero}b illustrates this relation.
Indeed, this momentum-mass relation suggests a possibility of the
$O(3,2)$ space-time symmetry~\cite{bk06jpa}.

\par
Within the framework of the Lorentz group, it is possible, by making
an excursion to  infinite momentum where the mass hyperbola coincides
with the light cone, to then come back to the desired point.  On the
other hand, the mass formula of Eq.(\ref{cocy70}) allows us to go there
directly.  The decoherence mechanism of the coherency matrix makes this
possible.

\section*{Acknowledgements}
I would like to thank Professor Peter Levai and the organizers of this
conference for allowing me to present a plenary talk.  This paper is
largely based on the papers I published in the past with Marilyn Noz
and Sibel Ba{\c s}kal.  I would like to thank them for their prolonged
collaborations.
\par
I am very grateful to Professor Eugene Wigner.  From 1985 to 1990, I went
to Princeton frequently to do physics under his guidance.  This is the
reason why I am frequently introduced as Wigner's youngest student, as
in the present conference.
\par
When I was a graduate student at Princeton (1958-61), I was afraid of
him.  My thesis advisor was Professor Sam Treiman.  I am not the first
student of Treiman to be interested in Wigner's 1939 paper on the internal
space-time symmetries of elementary particles.   Steven Weinberg was
Treiman's first student.  In 1964, he published a series of papers to make
his Wigner's paper useful to particle physics of that time.  In one of his
papers~\cite{wein64c}, he discussed massless particles and struggled with
the four-by-four matrix given in Eq.(\ref{e44}).  This matrix is contained
Wigner's original paper~\cite{wig39}.

\begin{figure}
\centerline{\includegraphics[scale=0.8]{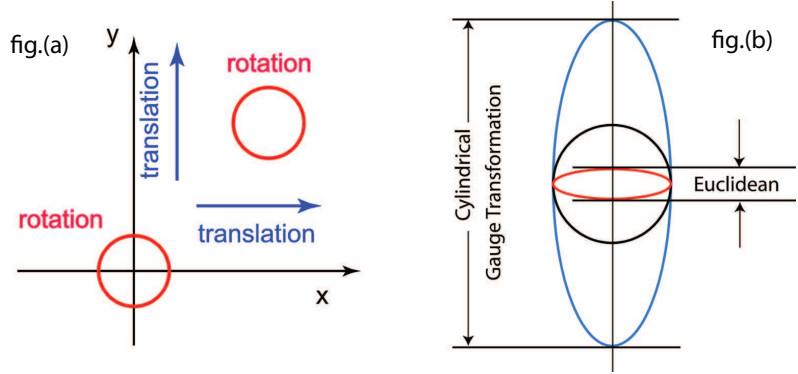}}
\caption{The geometry of the $E(2)$-like little group for massless
particles. In terms of There are one rotational and two translations
degrees of freedom in the two-dimensional Euclidean group, as shown
in fig.(a).  However, there is only one gauge degree of freedom.
Is it possible to collapse these two translational degrees into one?
The answer is that there is a cylindrical group isomorphic to
$E(2)$ where both the two translations in the $E(2)$ plane collapse
into one translation along the direction perpendicular to the plane,
as illustrated in fig.(b).}\label{eucl22}
\end{figure}

\par

Together with my younger colleagues, namely D. Han and D. Son, we were
able to interpret what Weinberg wanted to say as summarized in
Fig.~\ref{eucl22}a~\cite{hks82aa}.  According to this figure, the photon
spin corresponds the rotational on a two-dimensional Euclidean space, whose
Cartesian coordinates correspond to the gauge degrees of freedom.
\par

I explained this aspect of his $E(2)$-like little group to Professor Wigner.
He became very happy to hear that the translation-like variables correspond
to gauge transformations.  On the other hand, he noted that there is only
one gauge degree of freedom, while there are two translational degrees of
freedom.
\par
After some hard work, we came to the conclusion that the $E(2)$ symmetry can
be derived from a tangential plane on the north pole of the sphere.  For the
same sphere, we can consider a cylinder tangential to the equatorial belt of
the same sphere.  Equivalently, the sphere can be contracted or expanded
along the $z$ direction as shown in Fig.~\ref{eucl22}b~\cite{kiwi90jmp}.
The deformation takes place when the particle $(momentum/mass)$ changes from
zero to infinity.
\par
The two translational degrees collapse into one translation along the $z$
direction.  This cylinder can rotate around the same axis.  This rotational
degree corresponds to the photon spin. I had a pleasure of publishing this
paper with Wigner in the Journal of Mathematical Physics~\cite{kiwi90jmp}.
The editor of this journal was Lawrence Biedenharn at that time.  He told
me he was very happy to publish this paper in his journal.

\end{document}